\documentclass{PoS}

\usepackage{amsmath}
\usepackage{lineno}

\newcommand{\Xpr}{X^\prime_{2/3}}

\title{Interpretation of vector-like quark searches: the case of a heavy gluon in composite Higgs models and vector-like quarks}

\ShortTitle{Interpretation of vector-like quark searches}

\author{\speaker{Juan Pedro Araque} $^a$\thanks{J.P. Araque is supported by FEDER, COMPETE-QREN and FCT (project CERN/FIS-NUC/0005/2015) and grant SFRH/BD/52002/2012.} , Jos\'{e} Santiago$^b$\thanks{J. Santiago is supported by MINECO, under grant
numbers FPA2010-17915 and FPA2013-47836-C3-2-P, by the European Commission through the contract PITN-GA-2012-316704 (HIGGSTOOLS) and by Junta de Andaluc\'{\i}a grants FQM 101 and FQM 6552.} , Nuno Filipe Castro$^{a,c}$\thanks{N.F. Castro is supported by FEDER, COMPETE-QREN and FCT (project CERN/FIS-NUC/0005/2015) and contract IF/00050/2013.}\\

        a) Laborat\'orio de Instrumenta\c c\~ ao e F\'\i sica Experimental
de Part\'\i culas, Departamento de F\'\i sica da Universidade do Minho,
Campus de Gualtar, 4710-057 Braga, Portugal\\
b) CAFPE and Departamento de F\'\i sica Te\'{o}rica y del Cosmos, Universidad de Granada, E-18071 Granada, Spain\\
c) Departamento de F\'\i sica e Astronomia, Faculdade de Ci\^{e}ncias. Universidade do Porto, 4169-007 Porto, Portugal\\

        E-mail: \email{jaraque@lip.pt}, \email{jsantiago@ugr.es}, \email{nfcastro@lip.pt}}

\abstract{Pair production of new vector-like quarks in $pp$ collisions is considered model independent as it is usually dominated by QCD production. Nonetheless, the presence of a new massive color octet (heavy gluon) in some composite Higgs models may modify the pair production rate of vector-like quarks. This scenario is considered and the possible differences between the usual QCD production and the production mediated via heavy gluons is studied. The sensitivity to these differences in the LHC (run-1 and run-2) are studied. No sizeable differences have been found, which suggests that the published experimental results can be easily reinterpreted by a simple cross-section scaling. This reinterpretation has been also done for the run-1 results published by the collaborations.}

\FullConference{8th International Workshop on Top Quark Physics, TOP2015\\
		14-18 September, 2015\\
		Ischia, Italy}

\begin{document}
%\linenumbers
\section{Introduction}
Vector-like quarks can be produced via pair and single production. While the single production is clearly model dependent since it depends on the electroweak couplings of the vector-like quarks with the gauge bosons, pair production is considered model independent given that it is mediated via QCD interactions. Nonetheless, some composite Higgs models might include the presence of a new heavy gluon which will contribute to the pair production mechanism along with QCD interactions. In this case, the production rate will be increased due to a new production channel but it might be the case that some kinematical differences are introduced as well. If such differences exist, new analyses need to be developed to be sensitive to composite Higgs models with heavy gluons. On the other hand, if a deviation from the Standard Model is observed, the kinematical differences can be used to identify if that excess is due to the presence of a new heavy gluon. In the study presented here we evaluate the impact of a heavy gluon in the pair production of vector-like quarks and assess the sensitivity of current published analyses at $8$~TeV and the expected sensitivity with earily run-2 data ($L=10\text{ fb}^{-1}$) at $13$~TeV.

\section{The model}

For this study a minimal composite Higgs model (MCHM) has been used, in particular the MCHM4$_5$ %~\cite{DeSimone:2012fs,Anastasiou:2009rv}
 based on the coset $SO(5)/SO(4)$~\cite{Araque:2015cna} (and references therein). In this model four vector-like quarks are introduced as two doublets in terms of $SU(2)_L\times U(1)_Y$: $(T,B)$ and $(X_{5/3},X_{2/3})$, with hypercharges 1/6 and 7/6 respectively. After EWSB, a linear combination of $T$ and $X_{2/3}$, that we denote as $X^\prime_{2/3}$, remains degenerate with $X_{5/3}$. The orthogonal combination, that we call $T^\prime$, and $B$ are somewhat heavier with a small mass splitting. In most of the parameter space the decay branching ratio (BR) into the $Z$ and $H$ bosons are 50\% for the $\Xpr$ and $T'$ quarks while the $X_{5/3}$ and $B$ quarks decay entirely through charged currents.

In this scenario, following the partial compositeness mechanism, a composite gluon is introduced. While the elementary gluon only couples with the elementary quarks, the composite gluon only couples with the composite quarks. The elementary-composite gluon system can be brought to the physical basis by a given rotation. After this rotation we have a massless color octet, the SM gluon $g_\mu$, and a heavy gluon, $G_\mu$, with mass $M_G$. In the following we will use $M_G = 3$~TeV and $M_{X'_{2/3}}=1$~TeV as our benchmark masses.

%\begin{equation}
%\begin{pmatrix}
%G^e_\mu \\ G^c_\mu 
%\end{pmatrix}
%=
%\begin{pmatrix}
%\cos \theta_3 & -\sin \theta_3 \\
%\sin \theta_3 & \cos \theta_3
%\end{pmatrix}
%\begin{pmatrix}
%g_\mu \\ G_\mu
%\end{pmatrix},
%\end{equation}
%where the ratio of couplings fixes the mixing angle $\tan\theta_3=\frac{g_e}{g_c}$. 
\section{Kinematical differences}
Using the model presented in the previous section, Monte Carlo simulated samples (using \texttt{MadGraph}) have been used to study the effect of a heavy gluon in the kinematical signature in the pair production of vector-like quarks. Samples have been generated for the center of mass energies of 8~TeV and 13~TeV and considering the pair of vector-like quarks produced via QCD only (QCD), via a heavy gluon only (HG) and considering both mechanism with the proper interference between them (QCD+HG). For the sake of simplicity we will focus only in the $Z$ boson decay channel of the vector-like $X'_{2/3}$ in order to study the possible kinematical differences. 

\begin{figure}[ht!]
\begin{center}
\includegraphics[width=0.45\textwidth]{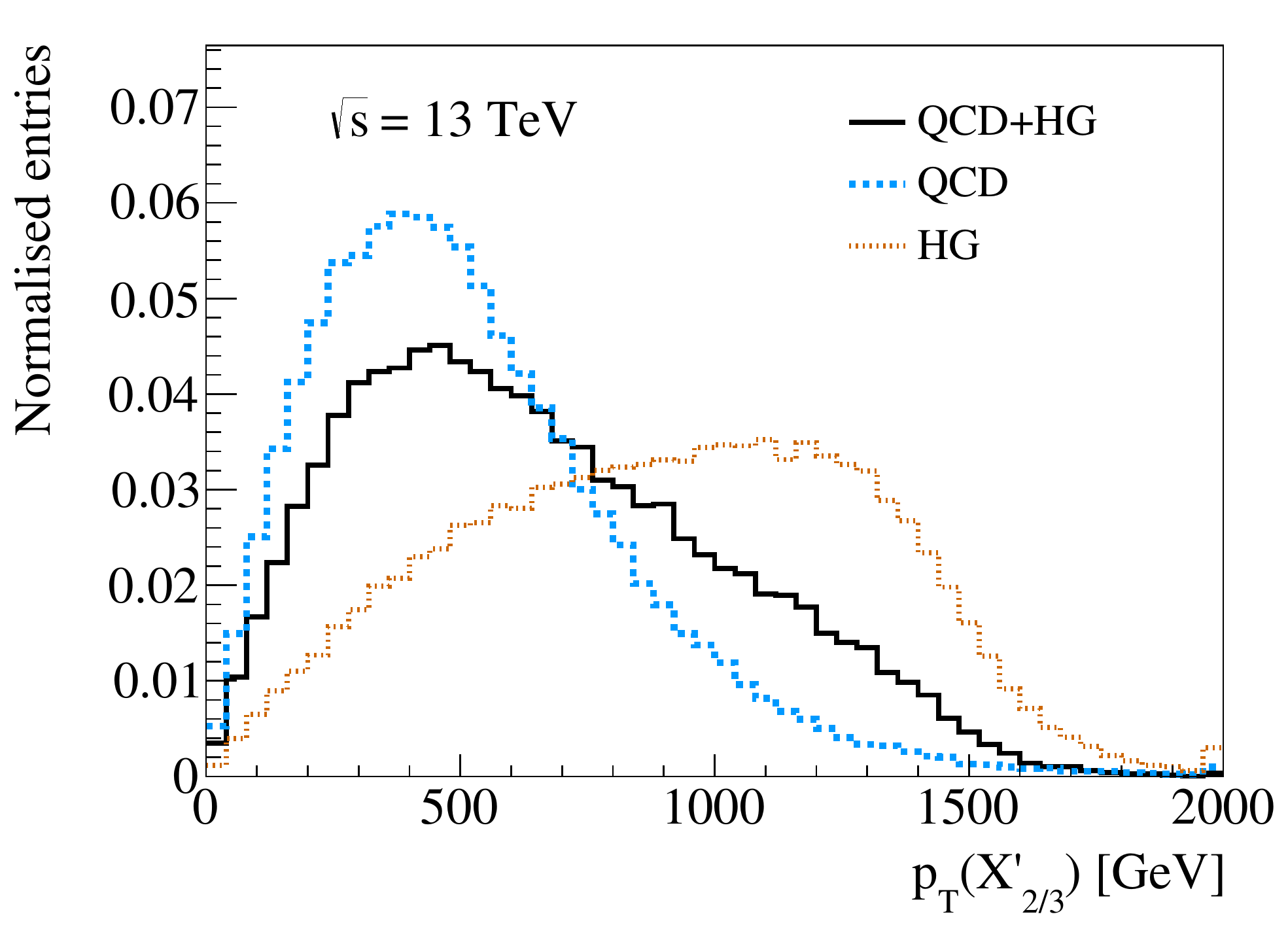}
\includegraphics[width=0.45\textwidth]{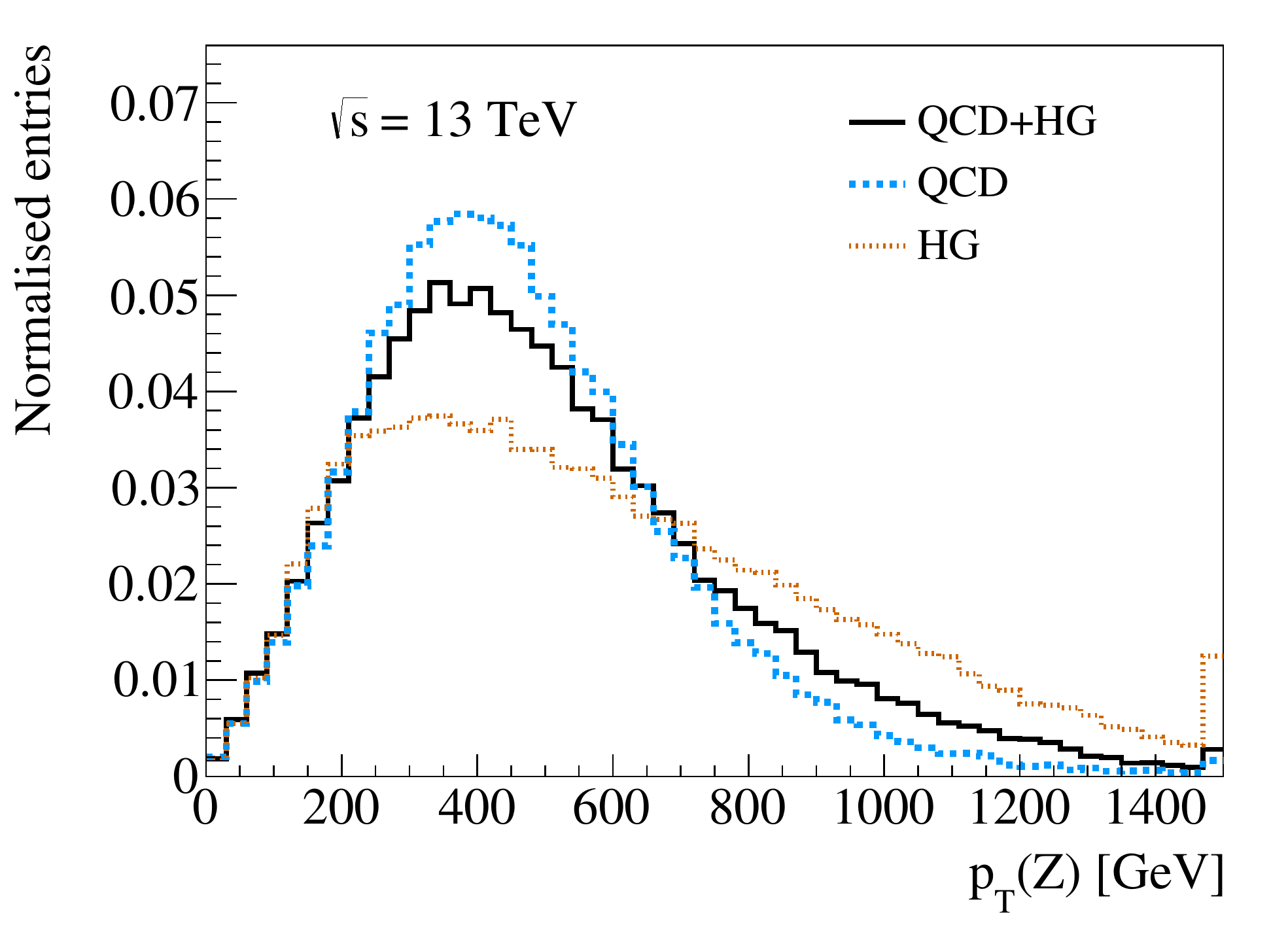}
\end{center}
\caption{$\Xpr$ $p_{\mathrm{T}}$ distribution (left) and $Z$ ${p_\mathrm{T}}$ distribution (right) at parton level at 13~TeV. Both distributions are normalized to unit area.}
\label{fig:kine}
\end{figure}

First, we study the kinematical differences at parton level. We found that the further in the decay chain from the heavy gluon the smaller the differences are between the different production mechanisms, as can be seen in Figure~\ref{fig:kine} where the transverse momentum of the $Z$ boson and the $X'_{2/3}$ are presented. Only 13~TeV distributions are shown as an example since they show bigger difference than the 8~TeV case. It can be seen that the effect of the heavy gluon is more noticeable in the $\Xpr$ distribution than in the $Z$ one. 

\begin{figure}[ht!]
\begin{center}
\includegraphics[width=0.45\textwidth]{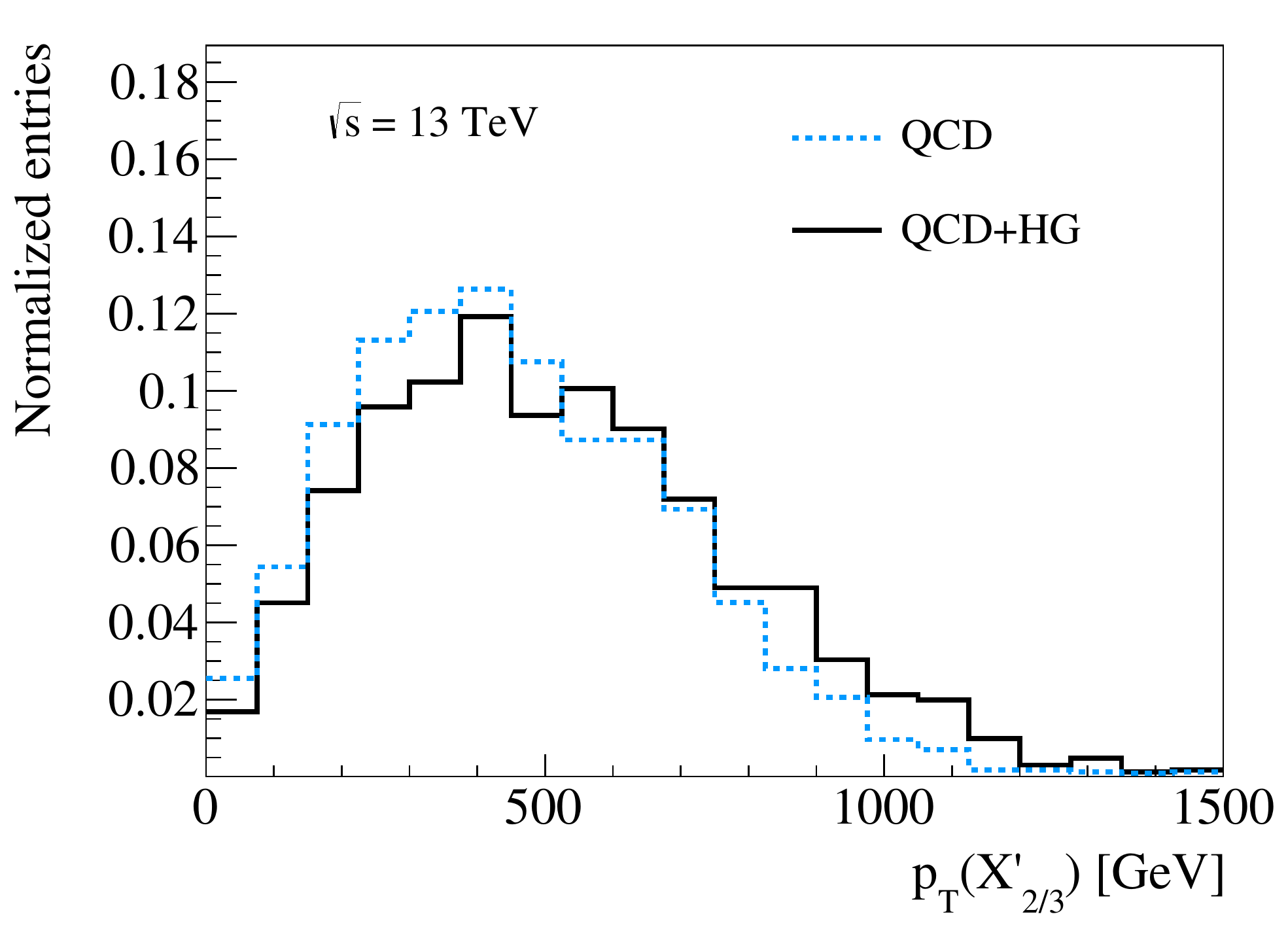}
\includegraphics[width=0.45\textwidth]{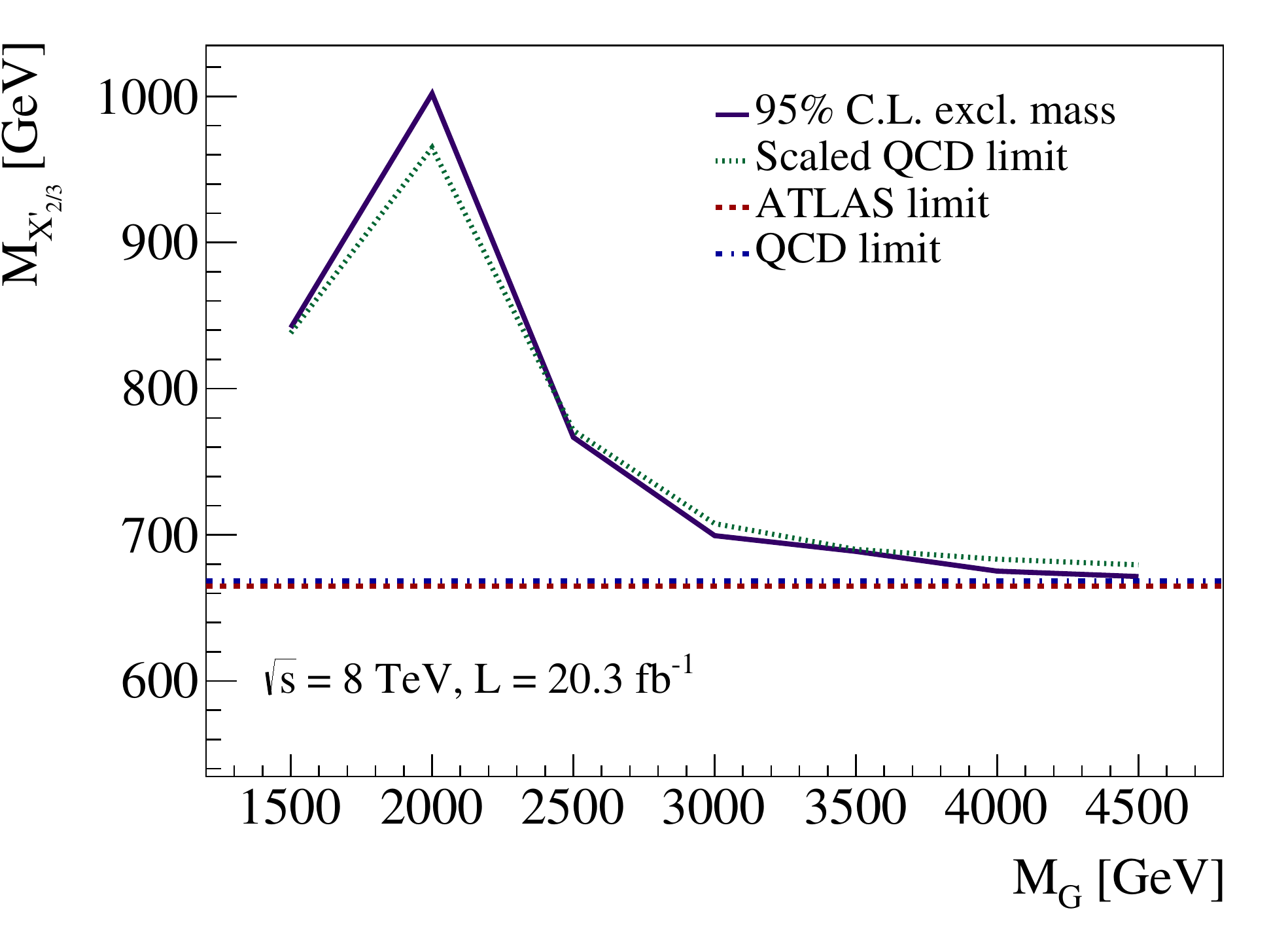}
\end{center}
\caption{$\Xpr$ $p_{\mathrm{T}}$ distribution (left) after hadronisation and detector simulation are taken into account. Expected lower mass limit after the neural network analysis at 8~TeV.}
\label{fig:delphes}
\end{figure}

After studying the heavy gluon effect at parton level, hadronisation and detector simulation are included (using \texttt{Pythia} and \texttt{Delphes}) to study wether or not the differences observed are kept in a more realistic scenario. After reconstructing the $\Xpr$ quark we observed that some kinematical differences are still maintained, for instance in the $p_{\rm T}(\Xpr)$ distribution as shown in Figure~\ref{fig:delphes} (left). In order to assess the effect of these differences in a realistic analysis a full recast of the the ATLAS analysis published in~\cite{Aad:2014efa} was performed, including the leading background sources, and no difference in the analysis sensitivity was found. Since the remaining differences are rather small after detector simulation, a more sophisticated analysis was performed using a neural network (implemented with \texttt{TMVA}) in order to try to exploit as much as posible the differences between both production mechanism. We used seven input variables for the neural network: Jet multiplicity, $p_{\rm T}(t)$, $p_{\rm T}(\Xpr)$, $p_{\rm T}(Z)$, $p_{\rm T}^{align}$, $H_{\rm T}$ and $M_{\Xpr}$. The $p_{\rm T}^{align}$ variable is sensitive to both the energy of the decay products of the vector-like quark and the angular separation between them. Even with the neural network analysis no significant differences were found between both production mechanism as can be seen in the lower mass limits shown in Figure~\ref{fig:delphes} (right) where the solid blue line represent the excluded mass when the heavy gluon production is included and the dashed green line represent the excluded mass when considering only QCD production with a scaled cross-section to match the QCD+HG production rate.

\section{Results}
\begin{figure}[ht!]
\begin{center}
\includegraphics[width=0.45\textwidth]{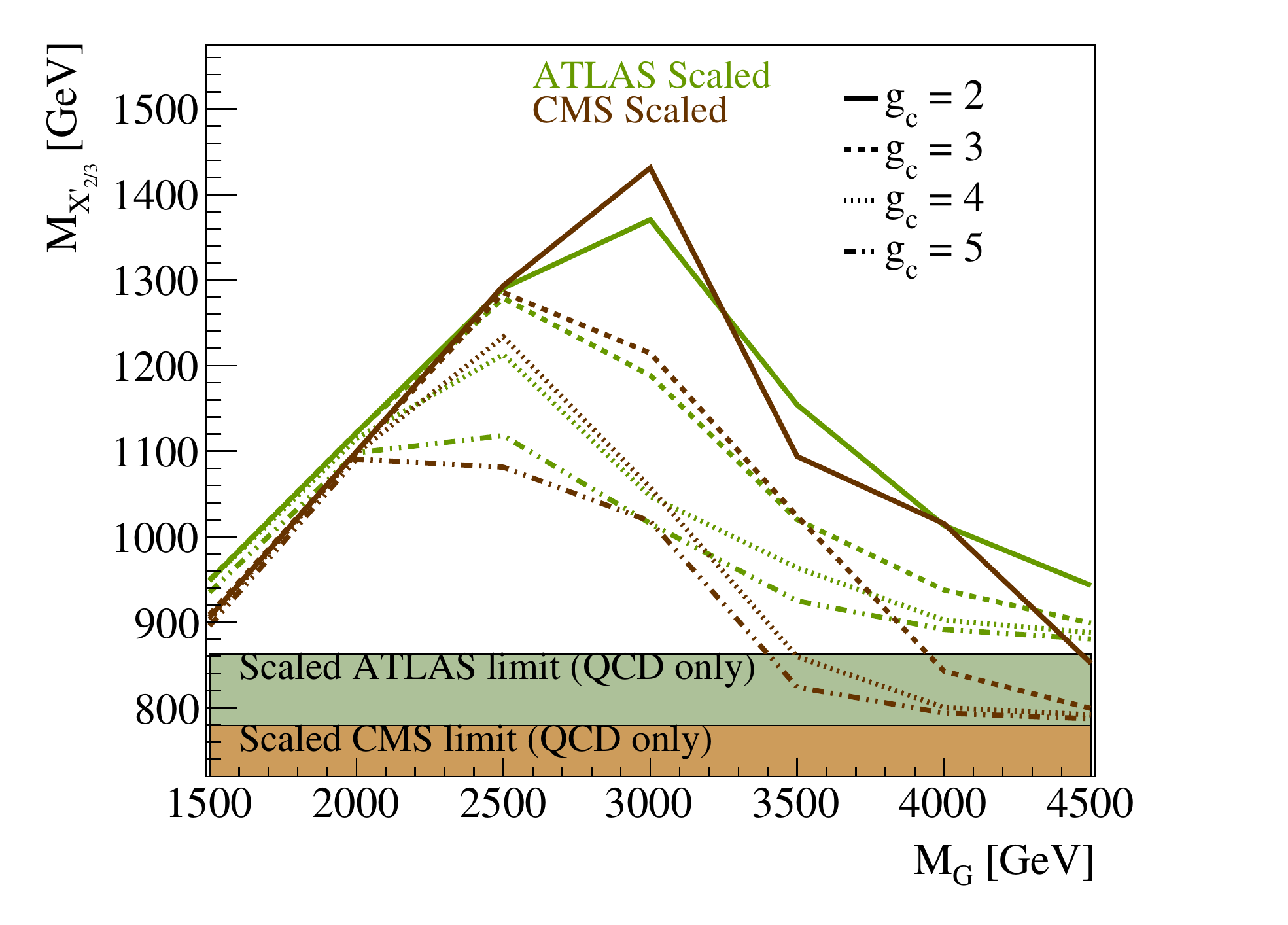}
\includegraphics[width=0.45\textwidth]{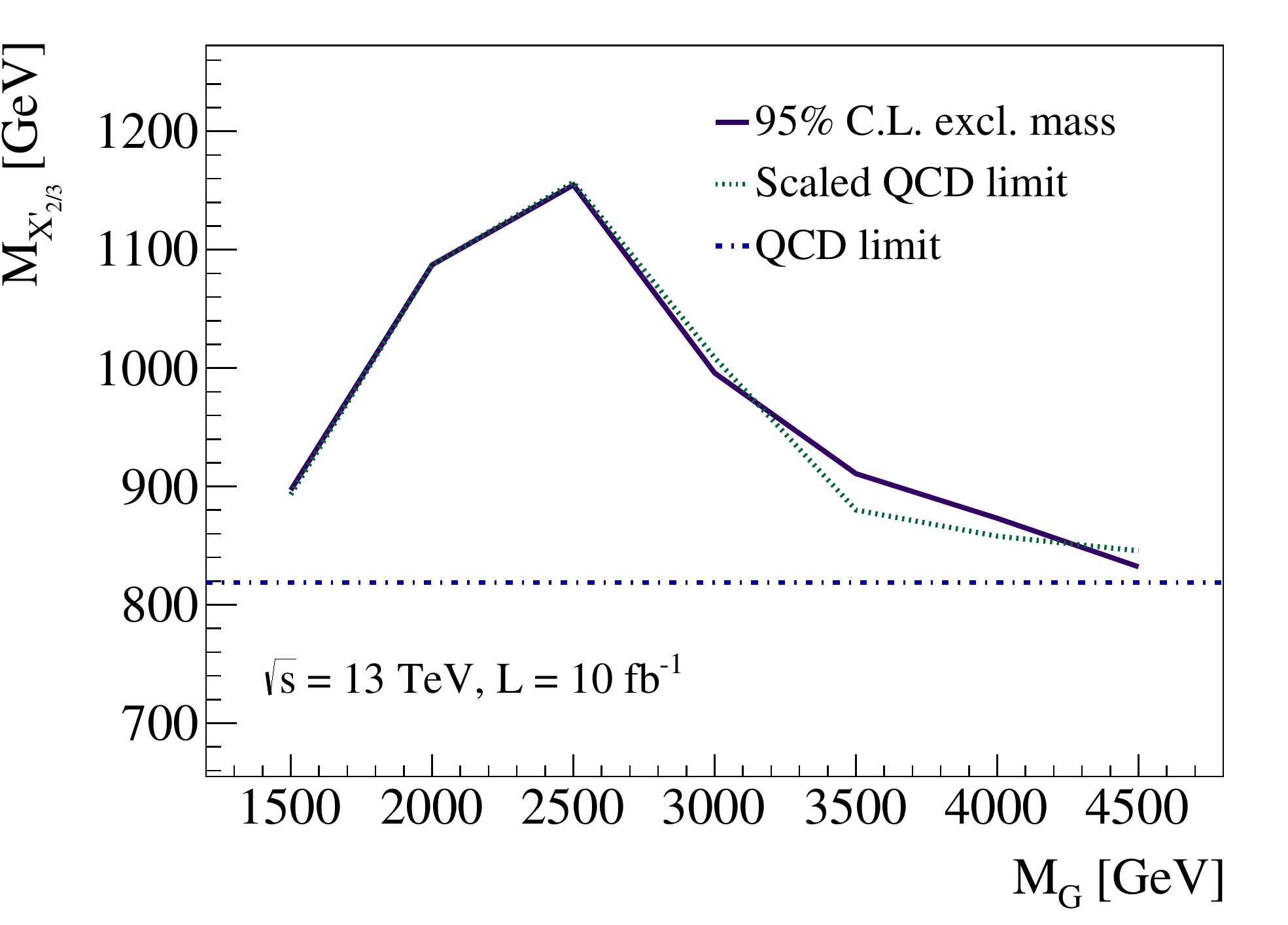}
\end{center}
\caption{95\% confidence level lower limits on the $M_G-M_{\Xpr}$ plane, derived from full run-1 published
data from ATLAS and CMS (left) and expected sensitivity for early run-2 data (right).}
\label{fig:res}
\end{figure}

In the previous analyses we saw that the presence of a heavy gluon does not have a significant impact on the pair production of vector-like quarks, other than an increase in the cross-section. In this scenario one could easily reinterpret the limits published by the collaborations on vector-like quark searches to set limits on the heavy gluon mass. This was done for the $\Xpr$ (shown in Figure~\ref{fig:res} (left)) and $X_{5/3}$ quarks. The expected sensitivity for early run-2 data ($L=10\text{ fb}^{-1}$) at 13~TeV was also studied using the neural network analysis. The results are shown in~\ref{fig:res} (right) with the same color scheme as before. As can be seen, even at 13~TeV no significant difference is expected due to the presence of a heavy gluon. The early run-2 limits on the mass of the VLQ are in the $\sim 820-1160$~GeV range (to be compared with the equivalent $\sim 660-1000$ GeV at $\sqrt{s}=8$ TeV). 

\end{document}